\documentclass[final,5p,times,twocolumn,nopreprintline]{elsarticle}
\usepackage{amsmath,slashed,booktabs,dsfont}
\usepackage{graphicx,graphics}
\usepackage{dcolumn}
\usepackage[hyperfootnotes=false]{hyperref}
\usepackage{xspace}
\usepackage{color}
\usepackage{balance}
\usepackage{multirow}
\usepackage{multicol}

\usepackage{fancyhdr}
\fancypagestyle{firstpage}{%
	
	\lhead{}
	\rhead{\small IPARCOS-UCM-23-139}
}
\newcommand{\diff}{\text{d}}

\newcommand{\mpi}{M_{\pi}}

\newcommand{\mN}{m_N}

\newcommand{\MeV}{\,\text{MeV}}
\newcommand{\GeV}{\,\text{GeV}}

\newcommand{\beq}{\begin{equation}}
\newcommand{\eeq}{\end{equation}}
\newcommand{\Wm}{W_\text{m}}
\renewcommand{\Re}{\text{Re}\,}
\renewcommand{\Im}{\text{Im}\,}

\allowdisplaybreaks[4]

\begin{document}

\renewcommand{\theequation}{\arabic{equation}}

\begin{frontmatter}
 
\title{Nucleon resonance parameters from Roy--Steiner equations}

\author[Bern]{Martin Hoferichter}
\author[Madrid]{Jacobo Ruiz de Elvira}
\author[Bonn]{Bastian Kubis}
\author[Bonn,Julich]{Ulf-G.\ Mei{\ss}ner}

\address[Bern]{Albert Einstein Center for Fundamental Physics, Institute for Theoretical Physics, University of Bern, Sidlerstrasse 5, 3012 Bern, Switzerland}
\address[Madrid]{Universidad Complutense de Madrid, Facultad de Ciencias F\'isicas,
Departamento de F\'isica Te\'orica and IPARCOS, Plaza de las Ciencias 1, 28040 Madrid, Spain}
\address[Bonn]{Helmholtz-Institut f\"ur Strahlen- und Kernphysik (Theorie) and
Bethe Center for Theoretical Physics, Universit\"at Bonn, 53115 Bonn, Germany}
\address[Julich]{Institute for Advanced Simulation (IAS-4),  Forschungszentrum J\"ulich, 52425  J\"ulich, Germany}

\begin{abstract}
 A reliable determination of the pole parameters and residues of nucleon resonances is notoriously challenging, given the required analytic continuation into the complex plane. We provide a comprehensive analysis of such resonance parameters accessible with Roy--Steiner equations for pion--nucleon scattering---a set of partial-wave dispersion relations that combines the constraints from analyticity, unitarity, and crossing symmetry---most prominently of the $\Delta(1232)$ resonance. Further, we study the Roper, $N(1440)$, resonance, which lies beyond the strict domain of validity, in comparison to Pad\'e approximants, comment on the role of subthreshold singularities in the $S$-wave, and determine the residues of the $f_0(500)$, $\rho(770)$, and $f_0(980)$ resonances in the $t$-channel process $\pi\pi\to\bar NN$. The latter allows us to test---for the first time fully model independently in terms of the respective residues---universality of the $\rho(770)$ couplings and  the Goldberger--Treiman relation expected if the scalars behaved as dilatons, in both cases revealing large deviations from the narrow-resonance limit.        
\end{abstract}

\end{frontmatter}

\thispagestyle{firstpage}

\section{Introduction}

Understanding the particle spectrum of QCD is notoriously challenging, maybe best exemplified by the decade-long controversy surrounding the very existence of the $f_0(500)$  and, even more so, the nature of this lowest-lying resonance in QCD~\cite{Pelaez:2015qba}.
In particular, a reliable determination of the pole parameters and residues requires controlling the analytic continuation far into the complex plane, a daunting task when only limited input in the physical region is available. In the case of the $f_0(500)$, a breakthrough was achieved in Ref.~\cite{Caprini:2005zr} by first solving $\pi\pi$ scattering in the physical region using Roy equations~\cite{Roy:1971tc,Ananthanarayan:2000ht,Colangelo:2001df}---a set of partial-wave dispersion relations (PWDRs) that implements all constraints from analyticity, unitarity, and crossing symmetry---and then using the same dispersion relations to perform the analytic continuation. Subsequently, different variants and refinements of such PWDRs were studied~\cite{Garcia-Martin:2011iqs,Caprini:2011ky}, further corroborating the $f_0(500)$ parameters and extending the analysis to the $\rho(770)$ and $f_0(980)$ resonances~\cite{Garcia-Martin:2011nna,Moussallam:2011zg}. Moreover, similar techniques have been applied to determine two-photon couplings~\cite{Hoferichter:2011wk,Moussallam:2011zg,Hoferichter:2019nlq,Danilkin:2019opj}, and to extend the analysis to higher energies~\cite{Pelaez:2022qby} (addressing the controversial case of the $f_0(1370)$), to $\pi K$ scattering~\cite{Buettiker:2003pp,Descotes-Genon:2006sdr,Pelaez:2020gnd,Danilkin:2020pak} (resolving the situation around the $K_0^*(700)$ resonance), and to $\pi\eta$ scattering~\cite{Lu:2020qeo} (determining the resonance parameters of the $a_0(980)$). Over the last two decades, the application of PWDRs has thus substantially advanced our understanding of the mesonic resonance spectrum, to the extent that previously controversial states are now well established in the PDG review~\cite{ParticleDataGroup:2022pth}.   

In comparison, similarly rigorous studies of nucleon resonances are scarce. A first step was taken in a program setting up and solving PWDRs for pion--nucleon ($\pi N$) scattering~\cite{Ditsche:2012fv,Hoferichter:2015hva} in the form of Roy--Steiner (RS) equations~\cite{Baacke:1970mi,Steiner:1970mh,Steiner:1971ms,Hite:1973pm}, which has led to a number of applications regarding the pion--nucleon $\sigma$-term~\cite{Hoferichter:2015dsa,Hoferichter:2016ocj,RuizdeElvira:2017stg,Hoferichter:2023ptl}, low-energy constants in chiral perturbation theory (ChPT)~\cite{Hoferichter:2015tha,Siemens:2016jwj}, and nucleon form factors~\cite{Hoferichter:2012wf,Hoferichter:2016duk,Hoferichter:2018zwu,Crivellin:2023ter},
see also Refs.~\cite{Lin:2021umz,Lin:2021xrc,Antognini:2022xqf}.
Since RS equations are built upon hyperbolic dispersion relations, they automatically couple $s$-channel, $\pi N\to\pi N$, and $t$-channel, $\pi\pi\to \bar N N$, partial waves, all of which need to be solved in a coupled system of integral equations~\cite{Hoferichter:2015hva}.
For the analytic continuation into the complex plane, this means that not only $s$-channel resonances, most prominently the $\Delta(1232)$, can be studied, but also the residues of resonances contributing to the $t$-channel process, such as the $f_0(500)$, $\rho(770)$, and $f_0(980)$, are accessible. In this work, we provide a comprehensive study of all nucleon resonance parameters that can be captured by analytically continuing the system of RS equations in either Mandelstam variable.  

 \begin{figure*}[t]
\includegraphics[width=0.49\linewidth]{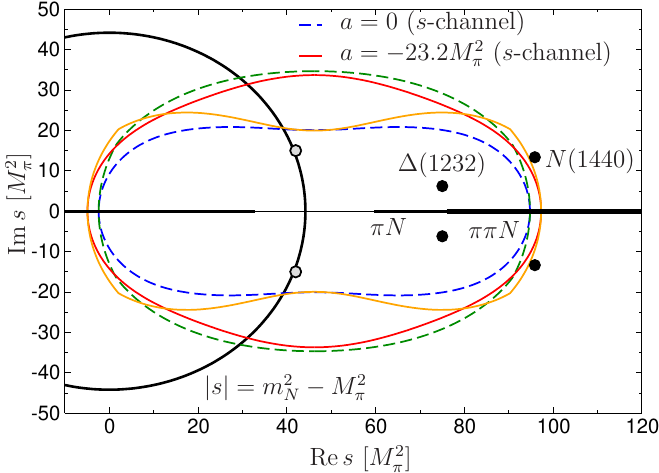}
\includegraphics[width=0.49\linewidth]{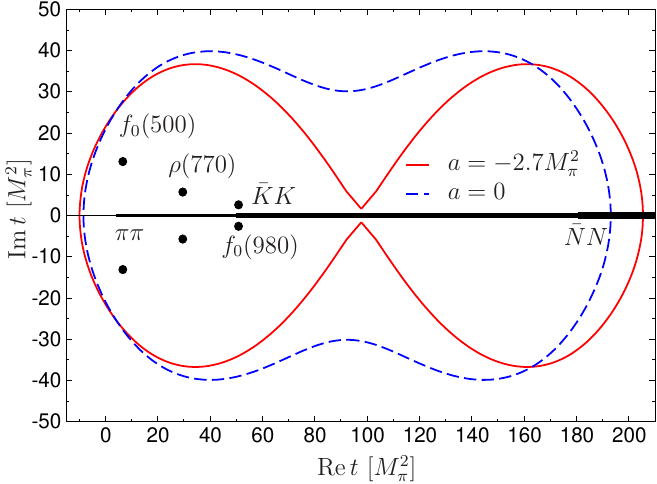}
\caption{Complex domain of validity for the $s$-channel (left) and $t$-channel (right) RS equations. The solid lines correspond to the optimized hyperbola parameters from Ref.~\cite{Ditsche:2012fv}, the dashed ones to the case $a=0$. Red and blue (green and orange) lines denote the constraints from the $s$- ($t$-) channel Lehmann ellipse, and the black lines the various cuts of the partial waves. Finally, the black dots indicate the (approximate) position of the known resonances, and the gray circles a subthreshold singularity in the $S$-wave that sits very close to the circular cut. }
\label{fig:validity}
\end{figure*}

In the $s$-channel, the most prominent resonance occurs in the isospin-$3/2$  $P$-wave, the well-established $\Delta(1232)$. However, even in this case, the uncertainty of mass and width of the resonance amounts to a few MeV~\cite{ParticleDataGroup:2022pth}, motivating a more rigorous determination based on the RS approach. Moreover, it has been found that a substantial part of the uncertainty in predictions of the neutrino--nucleus cross section at next-generation neutrino-oscillation experiments derives from $N\to\Delta$ form factors~\cite{Simons:2022ltq,Hernandez:2010bx,Ruso:2022qes}, and consolidating the $\Delta$ pole parameters thus constitutes an important first step for improvements, both from experiment and in lattice QCD. Second, the Roper, $N(1440)$, resonance lies beyond the strict range of applicability of the RS equations, but we can still obtain estimates for its pole parameters from Pad\'e approximants, after benchmarking the method for the $\Delta$ resonance. In addition, a Pad\'e-based extraction allows us to directly access the Riemann sheet closest to the physical region, which otherwise becomes a complicated task due to the interplay of $\pi N$ and $\pi\pi N$ cuts. 
 Finally, an $S$-wave subthreshold singularity deep in the complex plane was observed in Refs.~\cite{Wang:2017agd,Li:2021oou,Cao:2022zhn}, and we will comment on such a structure in our formalism.

In the $t$-channel, a rigorous determination of the various residues allows us to study to which extent the vector and tensor couplings $g_{\rho NN}^{(1)}$ and $g_{\rho NN}^{(2)}$ obtained from the analytic continuation of
the nucleon electromagnetic form factors comply with the expectations from a narrow-resonance approximation, improving upon previous determinations~\cite{Hohler:1974ht}, e.g., by profiting from modern experimental input on $\pi\pi$ and $\pi N$ scattering. Moreover, we can, for the first time, provide a reliable determination of the $S$-wave couplings $g_{SNN}$, for $S=f_0(500)$ or $S=f_0(980)$, defined model-independently as the residues at the respective poles, and study whether a Goldberger--Treiman relation
$F_S g_{SNN}=\mN$~\cite{Carruthers:1971vz,Goldberger:1958tr,Goldberger:1958vp,Nambu:1960xd,Gell-Mann:1960mvl} is fulfilled when $g_{SNN}$ and the decay constant $F_S$ are interpreted as (complex) residues of the pertinent amplitudes. Renewed interest in such considerations has been triggered recently in the context of one-boson-exchange models~\cite{Wu:2023uva} and dilaton physics~\cite{Crewther:2013vea,Cata:2018wzl,Zwicky:2023krx}, providing further motivation to clarify the $f_0(500)$ properties beyond the narrow-resonance picture.

\section{Roy--Steiner equations}

The complete system of RS equations decomposes into a set of PWDRs for each channel, e.g., for the $s$-channel partial waves
$f_{lI}^{I_s}$, where the index $I=\pm$ determines the total angular momentum $l\pm 1/2$ and $I_s=\{1/2,3/2\}$ the $s$-channel isospin,  according to
\begin{align}
\label{sRSpwhdr}
f^{I_s}_{lI}(W)&=N^{I_s}_{lI}(W)
+\frac{1}{\pi}\int\limits^{\infty}_{t_\pi}\diff t'\sum_{JI'}
G^{I_s}_{lJII'}(W,t')\,\Im f^J_{I'}(t')\notag\\
&+\frac{1}{\pi}\int\limits^{\infty}_{W_+}\diff W'\sum_{l'I_s'I'}K_{ll'II'}^{I_sI'_s}(W,W')\,\Im f^{I'_s}_{l'I'}(W'),
\end{align}
with $\mN$, $\mpi$  nucleon and pion masses, $s$, $t$ Mandelstam variables, $W=\sqrt{s}$, and thresholds $W_+=\mN+\mpi$, $t_\pi=4\mpi^2$.  $N^{I_s}_{l\pm}(W)$ refers to the nucleon Born terms, $f^J_\pm(t)$ to the $t$-channel partial waves, and the kernel functions $K_{ll'II'}^{I_sI'_s}$, $G^{I_s}_{lJII'}$ emerge from the partial-wave projection of the underlying hyperbolic dispersion relations~\cite{Hite:1973pm,Frazer:1960zz,Frazer:1960zza}. More details on the conventions for the partial waves are provided in~\ref{app:schannel} and~\ref{app:tchannel}, and for the full details of the RS system we refer to Ref.~\cite{Hoferichter:2015hva}. Crucially, the analytic continuation of the functions $N^{I_s}_{l\pm}$, $K_{ll'II'}^{I_sI'_s}$, and $G^{I_s}_{lJII'}$ in $W$ is known (and similarly in $t$ for the $t$-channel equations), in such a way that the RS formalism allows for a robust evaluation of the partial waves in the complex plane once their imaginary parts are determined. The solution of the coupled system of PWDRs is described in great detail in Refs.~\cite{Ditsche:2012fv,Hoferichter:2012wf,Hoferichter:2015hva}, based on the following input: (i) $\pi\pi$ and $\pi\pi\to\bar K K$ amplitudes for the $t$-channel equations~\cite{Buettiker:2003pp,Garcia-Martin:2011iqs,Caprini:2011ky}; (ii) $\pi N$ phase shifts and elasticities for the high-energy part of $s$-channel partial waves as well as partial waves beyond $P$-waves~\cite{Koch:1980ay,Hohler:1984ux,Arndt:2006bf,Workman:2012hx}; (iii) $\pi N$ scattering lengths from pionic atoms~\cite{Strauch:2010vu,Hennebach:2014lsa,Hirtl:2021zqf,Baru:2010xn,Baru:2011bw}. As output, the solution gives the $s$-channel partial waves with $l\leq 1$ below the matching point $\Wm=1.38\GeV$ and the $t$-channel ones with $J\leq 2$ up to the two-nucleon threshold.

These kinematic ranges are determined by the domain of validity of the RS system, which follows from the consideration of the large Lehmann ellipse~\cite{Lehmann:1958ita} and Mandelstam double-spectral functions~\cite{Mandelstam:1958xc}. In particular, hyperbolic dispersion relations are constructed along trajectories of constant $(s-a)(u-a)\equiv b$, where the hyperbola parameter $a$ can be chosen to maximize the domain of validity, leading to $a=-23.2\mpi^2$ and $a=-2.7\mpi^2$ for the $s$- and $t$-channel equations, respectively~\cite{Ditsche:2012fv}. Here, we need to extend this analysis into the complex plane, leading to the results shown in Fig.~\ref{fig:validity}. The argument follows closely the well-known strategy to construct the largest Lehmann ellipse in $b$ that does not reach into the double-spectral regions, see, e.g., Refs.~\cite{Caprini:2005zr,Descotes-Genon:2006sdr,Hoferichter:2011wk,Ditsche:2012fv,Hoferichter:2012pm}. We find that both for $a=0$ and for the optimized values of $a$ all interesting resonances are safely contained within the domain of validity,\footnote{We disagree with the findings of Ref.~\cite{Cao:2022zhn} that the $s$-channel RS equations can only be analytically continued for $a\geq -2.59\mpi^2$. This conclusion is drawn from an incorrect parameterization of the respective Lehmann ellipse.} the exception being the Roper $N(1440)$ just beyond the allowed region. However, in this case, another complication concerns the interplay with the $\pi\pi N$ cut, due to which this pole is, in any case, not directly accessible from the analytic continuation of the RS equations. For the analytic continuation of the $t$-channel equations, we find that the optimization along the real axis decreases the domain of validity in the complex plane, yet without affecting the low-energy resonances. In practice, we observe that the differences in the extracted resonance parameters between $a=0$ and the optimized values are negligible compared to other sources of uncertainty, and therefore continue to use the same values as in our previous work~\cite{Ditsche:2012fv,Hoferichter:2015hva}.

\begin{table}[t]
\renewcommand{\arraystretch}{1.3}
\centering
\scalebox{0.85}{
\begin{tabular}{crrrr}\toprule
& $M_R$ $[\text{MeV}]$ & $\Gamma_R$ $[\text{MeV}]$ & $|r|$ $[\text{MeV}]$ & $\delta$ $[^\circ]$\\\midrule
Ref.~\cite{Ronchen:2022hqk} & $1215(2)$ & $93(1)$ & $50(1)$ & $-39(1)$\\
Ref.~\cite{Svarc:2014zja} & $1211(1)(1)$ & $98(2)(1)$ & $50(1)(1)$ & $-46(1)(1)$\\ 
Ref.~\cite{Anisovich:2011fc} & $1210.5(1.0)$ & $99(2)$ & $51.6(6)$ & $-46(1)$\\
Ref.~\cite{Cutkosky:1980rh} & $1210(1)$ & $100(2)$ & $53(2)$ & $-47(1)$\\\midrule
Ref.~\cite{ParticleDataGroup:2022pth} & $[1209,1211]$ & $[98,102]$ & $[49,52]$ & $[-48,-45]$\\\midrule
This work & $1209.5(1.1)$ & $98.5(1.2)$ & $51.3(9)$ & $-47.4(4)$\\
\bottomrule
\end{tabular}}
\caption{Pole parameters of the $\Delta(1232)$, in the conventions of Ref.~\cite{ParticleDataGroup:2022pth}. The ranges quoted by Ref.~\cite{ParticleDataGroup:2022pth} are based upon Refs.~\cite{Ronchen:2022hqk,Svarc:2014zja,Anisovich:2011fc,Cutkosky:1980rh}.}
\label{tab:Delta_parameters}
\end{table}

\section{$\boldsymbol{\Delta(1232)}$}

The most prominent feature in $\pi N$ scattering is inarguably the $\Delta(1232)$ resonance. Despite its location close to the $\pi\pi N$ threshold, it is, to an excellent approximation, an elastic $\pi N$ resonance (in $f_{1+}^{3/2}$), and therefore ideally suited to determine its resonance position by analytic continuation of $\pi N$ RS equations. We obtain for mass and width
\begin{align}
\label{MGamma}
 M_R&=1209.49(83)(68)(15)(1)(11)\MeV,\notag\\ 
 \Gamma_R&=98.46(39)(81)(41)(8)(62)\MeV,
\end{align}
with a correlation $\rho_{M_R\Gamma_R}=0.65$.
The uncertainties arise from (i) flat directions in the space of subthreshold parameters, (ii) the phase shifts at the matching point, (iii) further input for $s$- and $t$-channel partial waves, (iv) the $\pi N$ scattering lengths, (v) the $\pi N$ coupling constant. In particular, we observe that the uncertainty in the $\pi N$ scattering lengths only plays a minor role, while the remaining systematic effects in the RS solution dominate, contrary to the situation in the case of the $\pi N$ $\sigma$-term. We can also determine the residue at the pole, see~\ref{app:schannel} for its precise definition, and find for modulus and phase
\begin{align}
\label{rdelta}
 |r|&=51.31(30)(57)(33)(6)(48)\MeV,\notag\\
 \delta&=-47.39(36)(1)(5)(2)(7)^\circ,
\end{align}
with a correlation $\rho_{|r|\delta}=0.11$. While our results are in agreement with previous determinations in the literature, see Table~\ref{tab:Delta_parameters}, the RS approach allows for unprecedented control over the uncertainties in each step of the calculation, 
both for the low-energy $\pi N$ amplitudes on the real axis and their analytic continuation. 

At this level of precision also the effects of isospin breaking need to be discussed. The pole parameters in Eqs.~\eqref{MGamma} and~\eqref{rdelta} refer to the isospin limit defined by the elastic reactions $\pi^\pm p\to\pi^\pm p$, motivated by the fact that these channels dominate the data base. Isospin-breaking corrections are taken into account accordingly, most notably in the scattering-length input~\cite{Gasser:2002am,Hoferichter:2009ez,Hoferichter:2009gn}. In consequence, the parameters for the $\Delta(1232)$ correspond to a weighted average of the $\Delta^{++}\simeq \pi^+ p$ and the $\Delta^0\simeq \pi^-p, \pi^0 n$ charge states. There have been studies of isospin-breaking effects in the $\Delta^{++}$--$\Delta^0$ system in the literature~\cite{Koch:1980ay,Pedroni:1978it,Abaev:1995cx,Bernicha:1995gg,Gridnev:2004mk}, including ChPT~\cite{Epelbaum:2007sq} and first results in lattice QCD~\cite{Tiburzi:2005na,CSSM:2019jmq,Romiti:2022rdk}, but a model-independent evaluation in terms of the respective pole parameters would require solving RS equations beyond the isospin limit. Such work has just begun for $\pi\pi$ scattering motivated by radiative corrections in $e^+e^-\to\pi^+\pi^-$~\cite{Colangelo:2022lzg,Monnard:2021pvm,Hoferichter:2023sli,Radinprep}, and would become even more challenging for $\pi N$ scattering.     

\begin{table}[t]
\renewcommand{\arraystretch}{1.3}
\centering
\scalebox{0.85}{
\begin{tabular}{crrrr}\toprule
& $M_R$ $[\text{MeV}]$ & $\Gamma_R$ $[\text{MeV}]$ & $|r|$ $[\text{MeV}]$ & $\delta$ $[^\circ]$\\\midrule
Ref.~\cite{Ronchen:2022hqk} & $1353(1)$ & $203(2)$ & $59(1)$ & $-104(2)$\\
Ref.~\cite{CBELSATAPS:2015kka} & $1369(3)$ & $189(5)$ & $49(3)$ & $-82(5)$\\
Ref.~\cite{Svarc:2014zja} & $1363(2)(2)$ & $180(4)(5)$ & $50(1)(2)$ & $-88(1)(2)$\\ 
Ref.~\cite{Cutkosky:1980rh} & $1375(30)$ & $180(40)$ & $52(5)$ & $-100(35)$\\\midrule
Ref.~\cite{ParticleDataGroup:2022pth} & $[1360,1380]$ & $[180,205]$ & $[50,60]$ & $[-100,-80]$\\\midrule
This work & $1374(3)(4)$ & $215(18)(8)$ & $58 (15)(17)$ & $-65(2)(11)$\\
\bottomrule
\end{tabular}}
\caption{Pole parameters of the $N(1440)$, where $r$ denotes the elastic residue. The ranges quoted by Ref.~\cite{ParticleDataGroup:2022pth} are based upon Refs.~\cite{Ronchen:2022hqk,Svarc:2014zja,Cutkosky:1980rh,CBELSATAPS:2015kka}. The uncertainties in our results refer to the $\pi N$ amplitude and the Pad\'e expansion, respectively, with correlation coefficients $\rho_{M_R\Gamma_R}=0.8$, $\rho_{|r|\delta}=-0.8$.}
\label{tab:Roper_parameters}
\end{table}

\section{$\boldsymbol{N(1440)}$}

Mass and width of the Roper are quoted in Ref.~\cite{ParticleDataGroup:2022pth} in terms of the ranges given in Table~\ref{tab:Roper_parameters}.
 As can be seen in Fig.~\ref{fig:validity}, the pole location lies close to the border of the strict range of validity, so that one might hope that useful information could still be obtained from the RS approach. Unfortunately, the analytic continuation of $f_{1-}^{1/2}$ via the $\pi N$ cut does not provide access to the pole closest to the physical region, which instead lies on the third sheet accessed via the $\pi\pi N$ cut. Accordingly, the pole obtained in analogy to the $\Delta(1232)$, $M_R=1473(35)\MeV$, $\Gamma_R=73(14)\MeV$, merely produces a reflection of the true Roper pole on the second sheet.  

Instead, we can use Pad\'e approximants to access the adjacent Riemann sheet~\cite{Montessus:1902}, at the expense of additional uncertainties from the choice of expansion point and degree of the Pad\'e series, see~\ref{app:Pade}. Following the strategy for error quantification from Ref.~\cite{Pelaez:2022qby} (see also Refs.~\cite{Masjuan:2013jha,Masjuan:2014psa,Caprini:2016uxy,Pelaez:2016klv,VonDetten:2021rax} for recent applications), we obtain the results summarized in Table~\ref{tab:Roper_parameters}. In comparison to previous work, we find a mass parameter  within the range from Ref.~\cite{ParticleDataGroup:2022pth}, while we see an indication for a slightly larger width. The elastic residue is also largely consistent, yet
affected by substantial Pad\'e uncertainties. In general, our uncertainties are significantly larger than the ones quoted in Refs.~\cite{Ronchen:2022hqk,Svarc:2014zja,CBELSATAPS:2015kka}, but already the spread among previous evaluations does cast doubt on the reliability of the uncertainty estimates. In this work, we have propagated both the uncertainties from the $\pi N$ amplitudes and the Pad\'e expansion, which should allow for a robust uncertainty quantification.    
As a cross-check on the method, we also evaluated the $\Delta(1232)$ parameters using Pad\'e approximants, yielding $M_R=1209.8(1.5)(0.1)\MeV$, $\Gamma_R=98.3(1.7)(0.2)\MeV$, $|r|=51.2(2.2)(0.1)\MeV$, $\delta=-46.8(2.2)(0.2)^\circ$, in excellent agreement with Eqs.~\eqref{MGamma} and~\eqref{rdelta}.

\section{Subthreshold singularities}

Subthreshold singularities in the $S$-wave $f_{0+}^{1/2}$ are known to occur when a given amplitude parameterization does not account for the full left-hand-cut structure~\cite{Doring:2009yv,Doring:2009uc}. More recently, such singularities were also reported in a simplified solution of RS equations~\cite{Cao:2022zhn}, which appears surprising since, by construction, the kernel functions in the PWDRs ensure the correct analytic behavior. Such a singularity is present in our solution as well, at $M_R=913.9(1.6)\MeV$, $\Gamma_R=337.7(6.2)\MeV$ (in good agreement with $M_R=918(3)\MeV$, $\Gamma_R=326(18)$ from Ref.~\cite{Cao:2022zhn}). Figure~\ref{fig:validity} shows that this singularity occurs close to the circular cut, and while numerically less pronounced than the $\Delta(1232)$ pole, we have checked that the singularity persists under the same parameter variations used before to quantify uncertainties in the RS solution. 

However, the physical interpretation of this singularity is far from obvious, as its position, far in the complex plane, casts doubts on its relevance for observables in the physical region. The situation can be contrasted with the $f_0(500)$ resonance, in which case a direct connection to $\pi\pi$ phase shifts can be established, e.g., by studying chiral trajectories~\cite{Pelaez:2015qba,Hanhart:2008mx,Pelaez:2010fj,Doring:2016bdr,Niehus:2020gmf}. Demonstrating such a connection in the $\pi N$ case will be challenging, in view of the large kinematic range to be spanned and less reliable effective-field-theory tools than in the mesonic case. For these reasons, we consider the interpretation of the above $S$-wave singularity an open question for the time being.    

\nocite{Hoferichter:2017ftn,Heuser:2024biq}

\begin{table}[t]
\renewcommand{\arraystretch}{1.3}
\centering
\scalebox{0.85}{
\begin{tabular}{crr}\toprule
& $f_0(500)$ & $f_0(980)$\\\midrule
$\sqrt{t_S}$ $[\text{MeV}]$ & 
$458(14)-279(10)i$ &
$1002(9)-23(8)i$\\
\multirow{2}{*}{$g_{S\pi\pi}$ $[\text{GeV}]$} & $0.95(19)-3.48(11)i$ & $0.40(26)-2.30(15)i$\\
& $3.61(13)e^{-1.30(5)i}$ & $2.33(18)e^{-1.40(10)i}$\\
\multirow{2}{*}{$F_S$ $[\text{MeV}]$} & $137(5)-58(5)i$ & $-150(26)-73(8)i$\\
& $149(6)e^{-0.40(3)i}$ & $167(25)e^{-2.69(8)i}$\\
\midrule  
& \multicolumn{2}{c}{$\rho(770)$}\\\midrule
$\sqrt{t_\rho}$ $[\text{MeV}]$ & \multicolumn{2}{c}{$762.5(1.7)-73.2(1.1)i$}\\
$g_{\rho\pi\pi}$ &\multicolumn{2}{c}{ $5.99(7)-0.54(12)i=6.01(8)e^{-0.09(2)i}$}\\
$g_{\rho\gamma}$ &\multicolumn{2}{c}{ $5.01(8)-0.11(9)i=5.01(7)e^{-0.02(2)i}$}\\
\bottomrule
\end{tabular}}
\caption{Resonance parameters for $f_0(500)$, $f_0(980)$, and $\rho(770)$. The pole positions and residues are taken from Ref.~\cite{Garcia-Martin:2011nna}, slightly adjusting the $S$-wave to be consistent with the input for the $\pi\pi/\bar K K$ $T$-matrix used in the RS solution~\cite{Hoferichter:2015hva}. The decay constants $F_S$ are defined in terms of the analytic continuation of the scalar form factor $\theta^\mu_\mu$~\cite{Donoghue:1990xh,Moussallam:2011zg} and the photon coupling $g_{\rho\gamma}$ parameterizes the residue of the electromagnetic form factor, see~\ref{app:tchannel}. In practice, we use the implementations from Refs.~\cite{Hoferichter:2012wf,Hoferichter:2017ftn} (extended to $\theta^\mu_\mu$ in the context of Refs.~\cite{Hoferichter:2016nvd,Hoferichter:2018acd}). The value of $g_{\rho\gamma}$ is updated according to the dispersive studies of $F_\pi^V$ in Refs.~\cite{Colangelo:2018mtw,Colangelo:2020lcg,Colangelo:2022prz}. The errors for $F_S$ and $g_{\rho\gamma}$ include uncertainties propagated from the respective form factor as well as the resonance parameters (including correlations).}
\label{tab:pipi_resonances}
\end{table}

\section{$\boldsymbol{t}$-channel residues}

The analytic continuation of the $t$-channel amplitudes is qualitatively different from the $s$-channel ones, in that resonance positions themselves are not accessible, but instead, the residues of known $\pi\pi$ resonances describing the coupling to $\bar N N$ can be determined. The resonance parameters, therefore, constitute input quantities for the current analysis but are derived from detailed dispersive analyses~\cite{Garcia-Martin:2011nna,Moussallam:2011zg,Hoferichter:2012wf,Hoferichter:2017ftn} and entail interesting applications themselves~\cite{Heuser:2024biq}. In particular, Table~\ref{tab:pipi_resonances} provides a comprehensive collection of the corresponding resonance parameters consistent with the hadronic amplitudes used for the solution of the $\pi N$ RS equations, see~\ref{app:tchannel} for precise definitions. Our results for the $t$-channel residues are collected in Table~\ref{tab:pipiNN_residues}.

Starting with the $\rho(770)$, the analytic continuation of the $\pi\pi\to\bar N N$ amplitudes allows us to quantify the departure of the vector and tensor couplings  $g_{\rho NN}^{(1)}$ and $g_{\rho NN}^{(2)}$ from narrow-resonance expectations. We find
\begin{align}
\label{grhoNN}
\frac{g_{\rho NN}^{(1)}}{g_{\rho\pi\pi}}&=0.50(12)+0.55(5)i=0.74(6)e^{0.82(15)i},\notag\\
\frac{g_{\rho NN}^{(2)}}{\kappa_vg_{\rho\pi\pi}}&=1.46(12)+0.54(3)i=1.55(11)e^{0.35(4)i},
\end{align}
where $\kappa_v=\kappa_p-\kappa_n=3.706$ is the isovector anomalous magnetic moment of the nucleon and both expressions become unity in the narrow-resonance limit. Compared to Ref.~\cite{Hohler:1974ht}, 
quoting $0.93(11)+0.45(5)i$, $1.69(20)+0.58(7)i$ for the two quantities (the uncertainty only refers to $g_{\rho\pi\pi}$), we observe similar imaginary parts, but significantly smaller real values. Part of the difference originates from $g_{\rho\pi\pi}^\text{\cite{Hohler:1974ht}}=5.7(3)$, the remainder from the analytic continuation of $f_\pm^1(t)$. A similar effect has also been observed in the case of the electromagnetic form factor of the pion $F_\pi^V$~\cite{Hoferichter:2017ftn}   
\beq
\label{grhog}
\frac{g_{\rho\gamma}}{g_{\rho\pi\pi}}=0.83(1)+0.06(2)i=0.83(1)e^{0.07(2)i},
\eeq
which in Ref.~\cite{Hohler:1974ht} is assumed to be unity, demonstrating the impact of modern data and robust techniques for the analytic continuation.
We conclude that both for the pion~\eqref{grhog} and the nucleon~\eqref{grhoNN} couplings to the $\rho(770)$ deviations from a narrow-width approximation are substantial, including sizable imaginary parts in the case of the nucleon.  This is also reflected in the tensor-to-vector coupling ratio,
$\kappa_\rho = \Re g_{\rho NN}^{(2)}/\Re g_{\rho NN}^{(1)} = 10.1(2.4)$ found here, for the first time with fully controlled and quantified uncertainties, compared to earlier determinations, $\kappa_\rho= 6.6$~\cite{Hohler:1974ht} or $\kappa_\rho = 6.1(2)$~\cite{Mergell:1995bf}. The relevance of such a large value for $\kappa_\rho$ in nuclear physics 
is discussed in Ref.~\cite{Brown:1994pq}.

\begin{table}[t]
\renewcommand{\arraystretch}{1.3}
\centering
\scalebox{0.85}{\begin{tabular}{crr}\toprule
& $f_0(500)$ & $f_0(980)$\\\midrule
\multirow{2}{*}{$g_{SNN}$} & $12.1(1.4)-13.9(5)i$ & $9.1(9)-2.9(5)i$\\
& $18.4(9)e^{-0.86(6)i}$ &
$9.6(9)e^{-0.31(6)i}$\\
\midrule  
& \multicolumn{2}{c}{$\rho(770)$}\\\midrule
$g_{\rho NN}^{(1)}$ &\multicolumn{2}{c}{$3.31(69)+2.99(36)i=4.46(36)e^{0.73(15)i}$}\\
$g_{\rho NN}^{(2)}$ &\multicolumn{2}{c}{$33.4(2.7)+9.0(1.0)i=34.6(2.4)e^{0.26(5)i}$}\\
\bottomrule
\end{tabular}}
\caption{Residues of the $t$-channel amplitudes $\pi\pi\to\bar N N$ describing the couplings of $f_0(500)$, $f_0(980)$, and $\rho(770)$  to $\bar N N$, see~\ref{app:tchannel} for precise definitions. The errors include uncertainties propagated from the resonance parameters as well as (i)--(v) for the $\pi\pi\to\bar N N$ partial waves (including correlations).}
\label{tab:pipiNN_residues}
\end{table}

Finally, for the $S$-waves, we can extract the coupling $g_{SNN}$, again defined model-independently in terms of the residue at the pole, see Table~\ref{tab:pipiNN_residues} for the results. Combined with the residues $F_S$ from Table~\ref{tab:pipi_resonances}, we can thus evaluate the combinations
\begin{align}
\label{GT}
 \frac{F_{f_0(500)}g_{f_0(500)NN}}{\mN}&=0.90(28)-2.78(20)i\notag\\
 &=2.93(24)e^{-1.26(9)i},\notag\\
 \frac{F_{f_0(980)}g_{f_0(980)NN}}{\mN}&=-1.69(27)-0.25(15)i\notag\\
 &=1.70(27)e^{-3.00(8)i},
\end{align}
which would be expected to become unity if the scalar could be interpreted as a dilaton, reflecting a generalized Goldberger--Treiman relation~\cite{Carruthers:1971vz}. Such interpretations have become of recent interest for the $f_0(500)$~\cite{Crewther:2013vea,Cata:2018wzl,Zwicky:2023krx}, but our result in Eq.~\eqref{GT} shows that as simple a relation as $F_S g_{SNN}=\mN$ is difficult to reconcile with its large width, leading to a sizable imaginary part in both $F_S$ and $g_{SNN}$ that does not cancel in the product. In fact, the real part of our result, $\Re g_{f_0(500) NN}=12.1(1.4)$, agrees well with $g_{f_0(500) NN}=12.2(2.3)\, [8.7(1.7)]$ from Ref.~\cite{Wu:2023uva}---based on 
matching SU(2)\, [SU(3)] ChPT and dispersion relations---and similarly for earlier work~\cite{Nagels:1979xh,Durso:1980vn,Machleidt:1987hj,Holzenkamp:1989tq,Reuber:1995vc,Ronchen:2012eg,Zhao:2013ffn,Liu:2018bkx}. The fact remains, however, that only the complex-valued residue is a physical observable.

\section{Conclusions}

In this work, we have provided a comprehensive analysis of nucleon resonance parameters that are accessible with Roy--Steiner equations. As key results, we obtained a precision determination of the pole parameters of the $\Delta(1232)$, a new evaluation of the Roper pole position via Pad\'e approximants, as well as the couplings of $f_0(500)$, $\rho(770)$, and $f_0(980)$ to $\bar N N$ defined model-independently in terms of the respective residues. Based on these results we could show that universality for the $\rho(770)$ couplings is strongly violated in nature, already in the mesonic case, but even more so for the nucleon. Moreover, we showed that the $f_0(500)$ does not fulfill a generalized Goldberger--Treiman relation when evaluating all ingredients in terms of complex residues, reaffirming that its large width impedes an interpretation as a dilaton field.  
 
\section*{Acknowledgments}

We thank Xiong-Hui Cao, Michael D\"oring, Feng-Kun Guo,  Qu-Zhi Li, Lewis C.~Tunstall, Han-Qing Zheng, and Roman Zwicky for useful discussions and correspondence. 
Financial support by the SNSF (Project No.\  PCEFP2\_181117), the DFG through the funds provided to the Sino--German Collaborative
Research Center TRR110 ``Symmetries and the Emergence of Structure in QCD''
(DFG Project-ID 196253076 -- TRR 110), 
the MKW NRW under the funding code NW21-024-A,
the Ram\'on y Cajal program (RYC2019-027605-I) of the Spanish MINECO, the Spanish Ministerio de Ciencia e Innovaci\'on (project  PID2022-136510NB-C31), and the European Research Council
(ERC) under the European Union's Horizon 2020 research and innovation program (ERC AdG EXOTIC, grant agreement
No.\ 101018170) is gratefully acknowledged.

\appendix

\section{Conventions for $\boldsymbol{s}$-channel amplitudes}
\label{app:schannel}

The $\pi N$ partial waves $f_{l\pm}^{I_s}(W)$ for isospin $I_s$ and angular momentum $j=l\pm 1/2$
are related to the $S$-matrix elements via~\cite{Frazer:1960zz}
\beq
S^{I_s}_{l\pm}(W)=1+2i|\mathbf{q}|f^{I_s}_{l\pm}(W),
\eeq
with center-of-mass momentum 
\beq
\label{q}
|\mathbf{q}|=\frac{\lambda^{1/2}(W^2,\mN^2,\mpi^2)}{2W},
\eeq
and $\lambda(a,b,c)=a^2+b^2+c^2-2(ab+ac+bc)$. They fulfill the unitarity relation
\begin{align}
 \Im f^{I_s}_{l\pm}(W)&=|\mathbf{q}|\big|f^{I_s}_{l\pm}(W)\big|^2\,\theta\big(W-W_+\big)\notag\\
 &+\frac{1-\big(\eta^{I_s}_{l\pm}(W)\big)^2}{4|\mathbf{q}|}\,\theta\big(W-W_\text{inel}\big),
\end{align}
where the elasticity parameters $\eta^{I_s}_{l\pm}(W)$ account for the inelastic effects above $W_\text{inel}=\mN+2\mpi$. The elastic $\pi N$ cut defines the analytic continuation onto the second sheet
\beq
\label{fl_AC}
f_{l\pm,\text{II}}^{I_s}(W)=\frac{f_{l\pm,\text{I}}^{I_s}(W)}{1-2i|\mathbf{q}|f_{l\pm,\text{I}}^{I_s}(W)},
\eeq
where we have already assumed $\Im W<0$ as required for the search for a resonance at 
\beq
W_R=M_R-i\frac{\Gamma_R}{2}. 
\eeq
In practice, the resonance position is therefore inferred from a null search of $1-2i|\mathbf{q}|f_{l\pm}^{I_s}(W)$ on the first sheet, evaluating the momentum according to Eq.~\eqref{q} for complex values of $W$. The residues are defined via~\cite{ParticleDataGroup:2022pth} 
\beq
f_{l\pm,\text{II}}^{I_s}(W)=\frac{r_{l\pm}^{I_s}}{|\mathbf{q}|(W_R-W)}.
\eeq

\section{Conventions for $\boldsymbol{t}$-channel amplitudes and form factors}
\label{app:tchannel}

We define the $\pi\pi$ partial waves $t_J^I(t)$ for isospin $I$ and angular momentum $J$ via the $S$-matrix elements
\beq
S_J^I(t)=1+2i\sigma_\pi(t) t_J^I(t),\qquad \sigma_\pi(t)=\sqrt{1-\frac{t_\pi}{t}},\qquad 
t_\pi=4\mpi^2,
\eeq
which implies the unitarity relation
\beq
\Im t_J^I(t)=\sigma_\pi(t)\big|t_J^I(t)\big|^2\theta(t-t_\pi).
\eeq
The residues are defined via~\cite{Garcia-Martin:2011nna}\footnote{Note that $g_{S\pi\pi}^\text{\cite{Moussallam:2011zg}}=\sqrt{2}g_{S\pi\pi}^\text{\cite{Garcia-Martin:2011nna}}$.} 
\beq
\label{tpipiII}
t_{0,\text{II}}^0(t)=\frac{g_{S\pi\pi}^2}{16\pi(t_S-t)},\qquad
t_{1,\text{II}}^1(t)=\frac{g_{\rho\pi\pi}^2(t-t_\pi)}{48\pi(t_\rho-t)},
\eeq
with resonance poles at
\beq
t_R=\bigg(M_R-i\frac{\Gamma_R}{2}\bigg)^2.
\eeq

Couplings to external currents are parameterized in terms of residues as well. For the vector current, the matrix element is defined as 
\beq
\langle\pi^\pm(p')|j^\mu(0)|\pi^\pm(p)\rangle = \pm(p+p')F_\pi^V(t),
\eeq
with $F_\pi^V(0)=1$ and unitarity relation
\beq
\Im F_\pi^V(t)=\sigma_\pi(t)\big[t_1^1(t)\big]^*F_\pi^V(t)\theta(t-t_\pi).
\eeq
Writing the form factor on the second sheet as 
\beq
\label{FpiVII}
F_{\pi,\text{II}}^V(t)=\frac{g_{\rho\pi\pi}}{g_{\rho\gamma}}\frac{t_\rho}{t_\rho-t},
\eeq
one has~\cite{Hoferichter:2017ftn}
\beq
\frac{1}{g_{\rho\gamma}g_{\rho\pi\pi}}=i\frac{\sigma_\pi^3(t_\rho)}{24\pi}F_{\pi,\text{I}}^V(t_\rho).
\eeq
We proceed analogously for the scalar current, i.e., starting from~\cite{Donoghue:1990xh}
\beq
\delta^{ab}F_{\pi}^\theta(t)=\langle\pi^a(p')|\theta^\mu_\mu|\pi^b(p)\rangle, 
\eeq
where $\theta^\mu_\mu$ denotes the trace of the energy-momentum tensor,
and the unitarity relation
\beq
\Im F_{\pi}^\theta(t)=\sigma_\pi(t)\big[t_0^0(t)\big]^*F_\pi^\theta(t)\theta(t-t_\pi),
\eeq
we define the form factor near the pole as 
\beq
\label{FpithetaII}
F^\theta_{\pi,\text{II}}(t)=\sqrt{\frac{2}{3}}\frac{F_S g_{S\pi\pi}t_S}{t_S-t},
\eeq
in such a way that in the narrow-width limit the residue $F_S$ coincides with the definition of the decay constant $\langle 0|\theta^\mu_\mu|S\rangle=F_S M_S^2$. Its general, complex value follows via analytic continuation
\beq
\frac{F_S}{g_{S\pi\pi}}=i\sqrt{\frac{3}{2}}\frac{\sigma_\pi(t_S)}{8\pi t_S}F^\theta_{\pi,\text{I}}(t_S). 
\eeq

The $\pi\pi\to\bar NN$ partial waves $f^J_\pm(t)$ are related to the $S$-matrix elements  according to~\cite{Frazer:1960zza,Hohler:1984ux,Ditsche:2012fv} 
\begin{align}
S^J_\pm(t)&=\frac{i}{c_J\sqrt{2}}\sqrt{\frac{p_t}{q_t}} F_\pm^J(t),\\
F_+^J(t)&=\frac{q_t}{p_t}(p_tq_t)^J\frac{2}{\sqrt{t}}f_+^J(t),\qquad 
F_-^J(t)=\frac{q_t}{p_t}(p_tq_t)^J f_-^J(t),\notag
\end{align}
where 
\beq
p_t=\sqrt{\frac{t}{4}-\mN^2},\qquad 
q_t=\sqrt{\frac{t}{4}-\mpi^2},
\eeq
the $\pm$ label refers to parallel or antiparallel antinucleon--nucleon helicities,
and the coefficients $c_J=1/\sqrt{6}$ ($1/2$) for $J$ even (odd) appear when expressing the $S$-matrix elements in isospin basis, corresponding to $I=0$ ($I=1$), respectively. The unitarity relation reads
\beq
\Im f_\pm^J(t)=\sigma_\pi(t)\big[t_J^I(t)\big]^*f_\pm^J(t)\theta(t-t_\pi).
\eeq
Starting with the scalar form factor, we have 
\begin{align}
\theta_N(t)&=\frac{1}{2\mN}\langle N(p')|\theta^\mu_\mu|N(p)\rangle,\notag\\
\Im \theta_N(t)&=\frac{\sigma_\pi(t)}{4\mN^2-t}\frac{3}{2}\big[F_\pi^\theta(t)\big]^*f_+^0(t)\theta(t-t_\pi),
\end{align}
and with residues parameterized as
\begin{align}
f_{+,\text{II}}^0(t)&=-\frac{p_t^2}{2\pi\sqrt{6}}\frac{g_{SNN}g_{S\pi\pi}}{t_S-t},\notag\\
\theta_{N,\text{II}}(t)&=\frac{F_S g_{SNN}t_S}{t_S-t},
\end{align}
we obtain 
\beq
\frac{g_{SNN}}{g_{S\pi\pi}}=i\sqrt{6}\frac{\sigma_\pi(t_S)}{4\mN^2-t_S}f_{+,\text{I}}^0(t_S).
\eeq
These parameterizations are again chosen in such a way that the narrow-width limit agrees with the standard Lagrangian definition, leading to the Goldberger--Treiman relation
$F_S g_{SNN}=\mN$~\cite{Carruthers:1971vz}. For simplicity, we have limited the presentation to $\pi\pi$ intermediate states, but for the numerical analysis we include $\bar K K$ contributions in the unitarity relations for all scalar amplitudes and form factors~\cite{Ditsche:2012fv,
Hoferichter:2012wf,Hoferichter:2015tha}, which is critical for a realistic description of the $f_0(980)$. 

Finally, for the $P$-wave we follow the conventions from Ref.~\cite{Hohler:1974ht} and define residues $g_{\rho NN}^{(i)}$ that in the narrow-width limit reduce to 
\beq
g_{\rho NN}^{(1)}=g_{\rho\pi\pi}=g_{\rho\gamma},\qquad 
g_{\rho NN}^{(2)}=(\kappa_p-\kappa_n)g_{\rho\pi\pi},
\eeq
referred to as vector and tensor coupling constants, respectively. The nucleon form factors are defined as 
\begin{align}
&\langle N(p')|j^\mu(0)|N(p)\rangle\notag\\
&=\bar u(p')\bigg[F_1^N(t)\gamma^\mu+\frac{i\sigma^{\mu\nu}q_\nu}{2\mN}F_2^N(t)\bigg]u(p),
\end{align}
with $q=p'-p$, and normalized to the charge and anomalous magnetic moment $\kappa_N$. Moreover, two-pion intermediate states contribute to the isovector combinations 
$F_i^v(t)=(F_i^p(t)-F_i^n(t))/2$, and the unitarity relations become simplest for the Sachs form factors
\begin{align}
G_E^N(t)&=F_1^N(t)+\frac{t}{4\mN^2}F_2^N(t),\notag\\
G_M^N(t)&=F_1^N(t)+F_2^N(t),
\end{align}
with
\begin{align}
 \Im G_E^v(t)&=\frac{\sigma_\pi(t) q_t^2}{2\mN}\big[F_\pi^V(t)\big]^*f_+^1(t)\theta(t-t_\pi),\notag\\
 \Im G_M^v(t)&=\frac{\sigma_\pi(t) q_t^2}{2\sqrt{2}}\big[F_\pi^V(t)\big]^*f_-^1(t)\theta(t-t_\pi).
\end{align}
The residues are parameterized according to 
\begin{align}
\label{GEMII}
 G_{E,\text{II}}^v(t)&=\frac{g_{\rho NN}^{(1)}+\frac{t_\rho}{4\mN^2}g_{\rho NN}^{(2)}}{2g_{\rho\gamma}}\frac{t_\rho}{t_\rho-t},\notag\\
  G_{M,\text{II}}^v(t)&=\frac{g_{\rho NN}^{(1)}+g_{\rho NN}^{(2)}}{2g_{\rho\gamma}}\frac{t_\rho}{t_\rho-t},\notag\\
  f_{+,\text{II}}^1(t)&=\frac{\mN g_{\rho\pi\pi}}{12\pi}\frac{g_{\rho NN}^{(1)}+\frac{t_\rho}{4\mN^2}g_{\rho NN}^{(2)}}{t_\rho-t},\notag\\
  f_{-,\text{II}}^1(t)&=\frac{\sqrt{2}\, g_{\rho\pi\pi}}{12\pi}\frac{g_{\rho NN}^{(1)}+g_{\rho NN}^{(2)}}{t_\rho-t},
\end{align}
and can be calculated from
\begin{align}
 \frac{g_{\rho NN}^{(1)}}{g_{\rho\pi\pi}}&=-2i\sigma_\pi(t_\rho)\frac{\mN q_t^2}{p_t^2}\bigg[f_{+,\text{I}}^1(t_\rho)-\frac{t_\rho}{4\sqrt{2}\,\mN}f_{-,\text{I}}^1(t_\rho)\bigg]\notag\\
 &=i\sqrt{2}\, \sigma_\pi(t_\rho)q_t^2\bigg[f_{-,\text{I}}^1(t_\rho)+\frac{\sqrt{2}\,\mN}{p_t^2}\Gamma^1_{\text{I}}(t_\rho)\bigg],\notag\\
 \frac{g_{\rho NN}^{(2)}}{g_{\rho\pi\pi}}&=2i\sigma_\pi(t_\rho)\frac{\mN q_t^2}{p_t^2}\bigg[f_{+,\text{I}}^1(t_\rho)-\frac{\mN}{\sqrt{2}}f_{-,\text{I}}^1(t_\rho)\bigg]\notag\\
 &=-2i\sigma_\pi(t_\rho)\frac{\mN q_t^2}{p_t^2}\Gamma^1_\text{I}(t_\rho),
\end{align}
where the linear combination 
\beq
\Gamma^J(t)=\mN\sqrt{\frac{J}{J+1}}f_-^J(t)-f_+^J(t)
\eeq
vanishes at $t=4\mN^2$. 

Finally, we comment on the parameterization of the dimensionful residues in Eqs.~\eqref{FpiVII}, \eqref{FpithetaII}, and \eqref{GEMII}, where, in analogy to the convention for the $\pi\pi$ residues~\eqref{tpipiII}, we kept the dependence on the full complex pole, while Refs.~\cite{Moussallam:2011zg,Hohler:1974ht} take $t_S\to M_S^2$, $t_\rho\to M_\rho^2$, respectively. 

\section{Pad\'e approximants}
\label{app:Pade}

A Pad\'e series for a function $f(s)$ is constructed in terms of approximants
\beq
P_M^N(s,s_0)=\frac{Q_N(s,s_0)}{R_M(s,s_0)},
\eeq
with polynomials $Q_N$ and $R_M$ of degree $N$ and $M$, in such a way that the first $N+M+1$ terms of the Taylor series around $s_0$ agree. De Montessus' theorem~\cite{Montessus:1902} guarantees that if $f(s)$ is regular inside a domain $\Omega$, except for poles with total multiplicity $M$, then the sequence $P_M^N(s)$ converges uniformly to $f(s)$ for $N\to\infty$ in any compact subset of $\Omega$ excluding the poles. Accordingly, Pad\'e approximants can be used to perform the analytic continuation to the adjacent Riemann sheet as long as 
the truncation error in $N$ can be controlled. 
In practice, we choose $M=1$ or $M=2$ (the additional pole serving as a means to mimic the effect of nearby singularities), study the convergence in $N$, and choose the point $s_0$ on the real axis to minimize the resulting statistical $\pi N$ and systematic Pad\'e error (determined as the difference between the highest orders $N$ and $N-1$ for which stable Pad\'e approximants can be constructed from the derivatives of $f(s)$).  
For more details on the numerical implementation and the estimate of the systematic effects, we refer to Ref.~\cite{Pelaez:2022qby}. 
The numerical results in the main text correspond to Pad\'e sequences with $M=1$, i.e., only one pole in the denominator. For the $N(1440)$, the minimum total error---including both truncation and $\pi N$ uncertainty effects---is found for the Pad\'e approximant $P^4_1$ at the optimal value $s_0=(1.35\GeV)^2$,  while in the $\Delta(1232)$ case, the most precise results are obtained for the approximant  $P^5_1$ at $s_0=(1.22\GeV)^2$. 
In addition, considering $P^N_2$ sequences, we found the Roper pole at $M_R=1362(5)(3)\MeV$ and $\Gamma_R=209(14)(5)\MeV$, hence fully consistent with the $P^4_1$ result. The second singularity is located around $\sqrt{s}\simeq (1.8 -0.3i)\GeV$ with large uncertainties.

\bibliographystyle{apsrev4-1_mod}
\balance
\biboptions{sort&compress}
\bibliography{ref}

\end{document}